\documentclass[twocolumn]{emulateapj}  
\usepackage{epstopdf}
\usepackage{ulem}
\usepackage{amssymb}
\usepackage{natbib}
\usepackage{times}
\usepackage{graphicx,bm,amssymb}
\usepackage{pstricks}

\usepackage{psfrag}
\usepackage[colorlinks=true,linkcolor=blue,citecolor=blue]{hyperref}
\newcommand{\be}{\begin{equation}}
\newcommand{\ee}{\end{equation}}
\newcommand{\bse}{\begin{subequations}}
\newcommand{\ese}{\end{subequations}}
\newcommand{\bary}{\begin{eqnarray}}
\newcommand{\eary}{\end{eqnarray}}

\shorttitle{}
\shortauthors{Fraija et al.}
\begin{document}
\title{Modeling the early afterglow in the short and hard GRB 090510}
\author{N. Fraija$^1$, W. H. Lee$^1$, P. Veres$^2$ and R. Barniol Duran$^3$}
\affil{$^1$Instituto de Astronom\'ia, Universidad Nacional Aut\'{o}noma de M\'{e}xico, Apdo. Postal 70-264, Cd. Universitaria, M\'{e}xico DF 04510\\
$^2$ Center for Space Plasma and Aeronomic Research (CSPAR), University of Alabama in Huntsville, Huntsville, AL 35899, USA\\
$^3$ Department of Physics and Astronomy, Purdue University, 525 Northwestern Avenue, West Lafayette, IN 47907, USA
}
\email{nifraija@astro.unam.mx, wlee@astro.unam.mx,  pv0004@uah.edu and rbarniol@purdue.edu}
\begin{abstract}
The bright, short and hard GRB 090510 was detected by all instruments aboard Fermi and Swift satellites. The multiwavelength observations of this burst presented similar features with the Fermi-LAT-detected gamma-ray bursts. In the framework of the external shock model of early afterglow, a leptonic scenario that evolves in a homogeneous medium is proposed to revisit GRB 090510 and explain the multiwavelength light curve observations presented in this burst. These observations are consistent with the evolution of a jet before and after the jet break. The long-lasting LAT, X-ray and optical fluxes are explained in the synchrotron emission from the adiabatic forward shock. Synchrotron self-Compton emission from the reverse shock is consistent with the bright LAT peak provided that progenitor environment is entrained with strong magnetic fields. It could provide compelling evidence of magnetic field amplification in the neutron star merger. 
\end{abstract}
\section{Introduction}
Gamma-ray bursts (GRBs) are intense and non repeating flashes with an extensive range of spectral and temporal properties. Based on the standard GRB durations, two dominant progenitor populations have been highly suggested, short ($T_{90}<$ 2 s) and long ($T_{90}>$ 2 s) GRBs. Despite the spectacular pace of discovery for long GRBs (lGRBs), the study of short GRBs (sGRBs) has proven much more challenging \citep{2014ARA&A..52...43B}. The most popular progenitor model associated with sGRBs is the merger of compact object binaries comprised of a neutron star binary (NS-NS) or a neutron star - black hole (NS-BH) \citep{1989Natur.340..126E, 1992ApJ...395L..83N, 2007NJPh....9...17L, 2004ApJ...608L...5L,2005ApJ...632..421L, 2007PhR...442..166N}.  LGRBs have been commonly associated  to core collapse of massive stars leading to supernovae (CCSNe) of type Ib,c and II  \citep{2006ARA&A..44..507W, 2012grbu.book..169H, 2003Natur.423..847H}. For lGRBs,  the strong gamma-ray polarization (\citet{2003Natur.423..415C, 2005A&A...439..245W} but see \citet{2004MNRAS.350.1288R} and more recently \citet{2011ApJ...743L..30Y}) and  a stronger magnetic field in reverse shock (RS) than in forward shock (FS) as a result of a full description of a dozen of lGRBs \citep{2005ApJ...628..315Z, 2003ApJ...595..950Z, 2007ApJ...655..391K, 2004A&A...424..477F, 2015ApJ...804..105F}, have supported the idea that the central engines of lGRBs could be endowed with primordial magnetic fields \citep{1992Natur.357..472U, 2000ApJ...537..810W}. In the framework of sGRBs, the question of what field strengths could be reached in a merger remnant before it collapses to a BH has so far remained unanswered, neither evidence of magnetization in the jet has been found.  Simulations have shown that during the merger the magnetic field strength can increase up to $10^{16}$ G.  The growth emerges from the transfer of hydrodynamical kinetic energy to electromagnetic energy via Kevin-Helmholtz instabilities and turbulent amplification \citep{2006Sci...312..719P, 2013ApJ...769L..29Z, 2009MNRAS.399L.164G,2010A&A...515A..30O}.\\
The bright, short and hard GRB 090510  was detected by Fermi and Swift satellites \citep{2010ApJ...709L.146D}.  During the first second after the GBM trigger at 00:22:59.97 UT, LAT detected 62  events with energies $>$ 100 MeV,  12 events with energies $>$ 1 GeV and one event (the most energetic photon) with $30.5^{+5.8}_{-2.6}$ GeV  at 0.829 s. The observations performed with the Very Large Telescope \citep{2009GCN..9353....1R, 2010A&A...516A..71M} provided a spectroscopic redshift of z=0.903 as measured [OII] and [H$\beta$] emission lines. Considering this redshift, the isotropic equivalent energy in the range of 10 keV - 30 GeV is $E\simeq (1.08\pm 0.06)\times 10^{53}$ erg. Refined analysis of BAT data found a duration of prompt emission T$_{90}=0.3\pm 0.1$ s \citep{2009GCN..9337....1U, 2010ApJ...709L.146D}.  The analysis of the prompt emission at latter times  ($\sim$ 0.3 s) showed a bright peak  in the LAT light curve (LC) above 100 MeV followed by a temporally extended LAT, X-ray and optical emission. These broadband observations by XRT and UVOT instruments on board Swift, were described by broken power laws with a break  at $\sim$ 1.5$\times 10^3$ s and LAT on board Fermi, was fitted with a simple power law extended from $\sim$ 1 s to $\sim\,200$ s. 
In the framework of  synchrotron radiation of FS,  a wide set of leptonic scenarios have been  explored to explain the high-energy photons ($\geq$ 100 MeV) present in GRB 090510  \citep{2011ApJ...730....1L,2011ApJ...733...22H,2010ApJ...718L..63P, 2009MNRAS.400L..75K, 2010MNRAS.409..226K,2010MNRAS.403..926G, 2010ApJ...720.1008C, 2010ApJ...709L.146D, 2010ApJ...724L.109R, 2010A&A...510L...7G}.  In particular,   \cite{2011ApJ...733...22H} described  not only LAT  but also X-ray and optical observations  from an early adiabatic FS emission in a model including a jet break.   The authors argued that the early emission ($t< 2$ s) was inconsistent with the standard RS. They explored the RS synchrotron emission in the thin-shell case,  when the flux peak at the crossing time was longer than the duration of the prompt emission ($t_{dec}> T_{90}$). They compared the RS synchrotron flux to FS synchrotron flux at 1 GeV, finding that this ratio was less than unity when the ejecta was not magnetized ($\sigma \ll1$ ) and  both the RS and FS synchrotron spectra were in the slow cooling regime.  After fitting the multiwavelength observations,  they found that the values of density, electron and magnetic microphysical parameters lie in the range  n $\sim\, 10^{-3} -10^{-6}\, {\rm cm^{-3}}$, $\epsilon_{e}\sim\,0.2 - 0.6$ and $\epsilon_{B}\sim 10^{-5} - 10^{-2}$, respectively. \\ 
Recently, \cite{2015ApJ...804..105F} and \cite{2016ApJ...818..190F} proposed an early afterglow model with in-homogenous density to explain the multiwavelength afterglow observations of GRB 110731A and GRB 130427A. As a result of good fits of the long-lasting LAT, X-ray and optical fluxes with synchrotron radiation from the adiabatic FS and  the bright LAT peak with synchrotron self-Compton (SSC) emission from RS in the thick-shell case, the author found that  central engine must be  entrained with strong magnetic fields. In this paper, we extend this early afterglow model to be evolved in a homogeneous medium and revisit GRB 090510 to describe the long-lasting emission and the bright peak flux present in the LAT LC.  
\section{Light Curves from external Shocks} 
In the following subsections we will show the LCs from FS and RS when electrons are accelerated and cooled down by synchrotron and Compton scattering emissions.
\subsection{LCs from FS emission}
The dynamics of the afterglow for a spherical ultrarelativistic and adiabatic shell propagating into a homogenous density is analyzed through the deceleration time, the energy of the shock, the bulk Lorentz factor and the radius at which the mass swept up \citep{1998ApJ...497L..17S, 1995ApJ...455L.143S, 2000ApJ...532..286K,1999A&AS..138..537S}.  Comparing  the time scale of synchrotron process with the deceleration time of the ejecta and the acceleration time of electrons, the synchrotron spectral breaks and the maximum flux are calculated in \cite{1998ApJ...497L..17S}.\\
Requiring the observed synchrotron spectrum for the fast- and slow-cooling regime and the synchrotron spectral breaks \citep{1998ApJ...497L..17S}, the temporal decay fluxes can be obtained as a function of energy. Then, the  LC obtained in the fast- and slow-cooling regime are
{\small
\begin{eqnarray}
\label{fcsyn_t}
F^{\rm syn}_{\nu,f}= \cases{ 
A^{\rm syn}_{\rm fl}\,t^{\frac{1}{6}}\, \left(E^{syn}_\gamma\right)^{\frac13},\hspace{1.3cm} E^{\rm syn}_\gamma<E^{\rm syn}_{\rm \gamma,c,f}, \cr
A^{\rm syn}_{\rm fm}\,t^{-\frac{1}{4}}\, \left(E^{syn}_\gamma\right)^{-\frac12},\,\,\,\, E^{\rm syn}_{\rm \gamma,c,f}<E^{\rm syn}_\gamma<E^{\rm syn}_{\rm \gamma,m,f}, \cr
A^{\rm syn}_{\rm fh}\,t^{-\frac{3p-2}{4}}\,\left(E^{syn}_\gamma\right)^{-\frac{p}{2}},\,\,\,\,E^{\rm syn}_{\rm \gamma,m,f}<E^{\rm syn}_\gamma<E^{\rm syn}_{\rm \gamma,max,f}\,, \cr
}
\end{eqnarray}
}
and
{\small
\begin{eqnarray}
\label{scsyn_t}
F^{\rm syn}_{\nu,f}=\cases{
A^{\rm syn}_{sl}t^{\frac12}\left(E^{\rm syn}_\gamma\right)^{\frac13},\hspace{1.3cm} E^{\rm syn}_\gamma<E^{\rm syn}_{\rm \gamma,m,f},\cr
A^{\rm syn}_{sm}t^{-\frac{3p-3}{4}}\left(E^{\rm syn}_\gamma\right)^{-\frac{p-1}{2}},E^{\rm syn}_{\rm \gamma,m,f}<E^{\rm syn}_\gamma<E^{syn}_{\rm \gamma,c,f},\,\,\,\,\,\cr
A^{\rm syn}_{sh}\,t^{-\frac{3p-2}{4}}\,\left(E^{\rm syn}_\gamma\right)^{-\frac{p}{2}},\,E^{\rm syn}_{\rm \gamma,c,f}<E^{syn}_\gamma<E^{\rm syn}_{\rm \gamma,max,f}\,, \cr
}
\end{eqnarray}
}
respectively. The coefficients $A^{\rm syn}_{\rm fl}$, $A^{\rm syn}_{\rm fm}$, $A^{\rm syn}_{\rm fh}$, $A^{\rm syn}_{\rm sl}$ and $A^{\rm syn}_{\rm sm}$ are given in appendix.\\
When the bulk Lorentz factor drops below the inverse of the jet opening angle $\Gamma_f\sim\theta_j^{-1}$, a break (at the jet-break time) in the LC  is observed because the surface which we can observe has grown larger than the surface that radiates.    This episode is evident in LCs, exhibiting a break in the afterglow observations \citep{1999ApJ...519L..17S}.   Therefore, afterglow emission can be affected by the evolution of the jet before and after  it slows down and spreads laterally.\\
Synchrotron photons can be up-scattered by Fermi-accelerated electrons \citep[e.g.][]{2014ApJ...787..168V, 2012ApJ...755..127S}.  Using the synchrotron spectral breaks \citep{1998ApJ...497L..17S}, the SSC spectral breaks and the maximum flux are computed in \cite{2001ApJ...548..787S}.\\
\subsection{LCs from RS emission}
We consider the thick-shell case for which  the shell is significantly decelerated by the RS (i.e. the bulk Lorentz factor is higher that the critical Lorentz factor $\Gamma_c$) \citep{2005ApJ...628..315Z,2012ApJ...751...33F}.   Numerical analysis performed by \cite{2004A&A...424..477F} revealed that for the particular value of the magnetization parameter $\sigma \simeq 1$ which is defined as the ratio of Poynting flux to matter energy flux  $\sigma =L_{pf}/L_{kn}\simeq \epsilon_{B,r}$  \citep{2004A&A...424..477F,2007ApJ...655..973K}, the shock crossing time becomes $t_d\simeq T_{90}/6$.\\
The synchrotron and SSC spectral breaks, and  fluxes are determined by the synchrotron spectral evolution between RS and FS for $\mathcal{R}_e=1$ \citep{2005ApJ...628..315Z, 2015ApJ...804..105F, 2016ApJ...818..190F}.\\
Before the RS crosses a thick shell, the hydrodynamic variables as a function of the observer time $t\propto R^2$ vary as \citep{2007ApJ...655..391K}: the bulk Lorentz factor $\Gamma_r \propto  t^{-1/4}$, the homogeneous density $n\propto  t^{-3/4}$, the pressure $p\propto t^{-1/2}$ and the total number of the shocked electrons $N_e\propto t$. Taking into account the evolution of the magnetic field  $B'_r\propto t^{-1/4}$, the minimum Lorentz factor $\gamma_{\rm e,m,r}\propto t^{1/4}$ and the cooling Lorentz factor $\gamma_{\rm e,c,r}\propto B'^{-2}_r\Gamma_r^{-1}t^{-1}\propto t^{1/4}$, then the synchrotron spectral breaks and the maximum flux for $t\,<\,T_{90}$ evolve as
{\small
\bary\label{Esyn_rtb}
E^{\rm syn}_{\rm \gamma,m,r}&\propto&\, {B'}_r\,\Gamma_r\, \gamma_{\rm e,m,r}^2\propto t^0,\nonumber\\
E^{\rm syn}_{\rm \gamma,c,r}&\propto&\, {B'}_r^{-3}\,\Gamma_r^{-1}\, t^{-2}\propto t^{-1},\\
F_{\rm \gamma,max,r}&\propto&\, N_e\, {B'}_r\, \Gamma_r\,\propto t^{1/2}\,\nonumber.
\eary
}
Considering the inverse Compton scattering equations and eq. (\ref{Esyn_rtb}), we obtain that the SSC LC  for fast cooling regime is
{\small
\begin{eqnarray}
\label{fcsyn_rbt}
F^{\rm ssc}_{\nu,r}\propto \cases{ 
t\,\left(E^{ssc}_{\rm \gamma,r}\right)^{\frac{1}{3}},\hspace{1.3cm} E^{\rm ssc}_\gamma<E^{\rm ssc}_{\rm \gamma,c,r}, \cr
t^{-\frac{1}{4}}\,\left(E^{\rm ssc}_{\rm \gamma,r}\right)^{-\frac{1}{2}},\,\,\,\, E^{\rm ssc}_{\rm \gamma,c,r}<E^{\rm ssc}_\gamma<E^{\rm ssc}_{\rm \gamma,m,r}, \cr
t^\frac{p-2}{4}\,\left(E^{\rm ssc}_{\rm \gamma,r}\right)^{-\frac{p}{2}},\,\,\,\,E^{\rm ssc}_{\rm \gamma,m,r}<E^{\rm ssc}_\gamma\,,\cr
}
\end{eqnarray}
}
and for slow cooling regime is
{\small
\begin{eqnarray}
\label{scsyn_rbt}
F^{\rm ssc}_{\nu,r}\propto\cases{
t^{\frac13}\,\left(E^{ssc}_{\rm \gamma,r}\right)^{\frac{1}{3}},\hspace{1.3cm} E^{\rm ssc}_\gamma<E^{\rm ssc}_{\rm \gamma,m,r},\cr
t^{\frac{p+1}{4}}\,\left(E^{ssc}_{\rm \gamma,r}\right)^{-\frac{p-1}{4}},E^{\rm ssc}_{\rm \gamma,m,r}<E^{\rm ssc}_\gamma<E^{\rm ssc}_{\rm \gamma,c,r},\,\,\,\,\,\cr
t^\frac{p-2}{4}\,\left(E^{ssc}_{\rm \gamma,r}\right)^{-\frac{p}{2}},\,E^{\rm ssc}_{\rm \gamma,c,r}<E^{\rm ssc}_\gamma. \cr
}
\end{eqnarray}
}
Once the RS has crossed the thick shell, the hydrodynamic variables as a function of the observer time $t\propto R^8$ vary as \citep{2007ApJ...655..391K}:  the bulk Lorentz factor $\Gamma_r \propto  t^{-7/16}$, the homogeneous density $n\propto  t^{-13/16}$, the pressure $p\propto t^{-13/12}$ and the total number of the shocked electrons $N_e\propto t^0$. Taking into account the evolution of the magnetic field  ${B'}_r\propto t^{-12/24}$, the minimum Lorentz factor $\gamma_{\rm e,m,r}\propto t^{-13/48}$ and the cooling Lorentz factor $\gamma_{\rm e,c,r}\propto {B'}_r^{-2}\Gamma_r^{-1}t^{-1}\propto t^{25/48}$, the synchrotron spectral breaks and the maximum flux for $t\,>\,T_{90}$ evolve as
{\small
\bary\label{Esyn_rta}
E^{\rm syn}_{\rm \gamma,m,r}&\propto&\, B_r\Gamma_r \gamma_{\rm e,m,r}^2\propto t^{-73/48},\nonumber\\
E^{\rm syn}_{\rm \gamma,c,r}&\propto&\, B_r^{-3}\Gamma_r^{-1} t^{-2}\propto t^{-1/16},\\
F_{\rm \gamma,max,r}&\propto&\, N_e B_r \Gamma_r\propto t^{-47/48}\,\nonumber.
\eary
}
Considering the inverse Compton scattering equations and eq. (\ref{Esyn_rta}), we get that the SSC LC  for fast cooling regime is
{\small
\begin{eqnarray}
\label{fcsyn_rat}
F^{\rm ssc}_{\nu,r}\propto \cases{ 
t^{-\frac53}\,\left(E^{ssc}_{\rm \gamma,r}\right)^{\frac{1}{3}},\hspace{1.3cm} E^{\rm ssc}_\gamma<E^{\rm ssc}_{\rm \gamma,c,r}, \cr
t^{-\frac{3}{4}}\,\left(E^{ssc}_{\rm \gamma,r}\right)^{-\frac{1}{2}} ,\,\,\,\, E^{\rm ssc}_{\rm \gamma,c,r}<E^{\rm ssc}_\gamma<E^{\rm ssc}_{\rm \gamma,m,r}, \cr
t^\frac{-4p+1}{4}\,\left(E^{ssc}_{\rm \gamma,r}\right)^{-\frac{p}{2}},\,\,\,\,E^{\rm ssc}_{\rm \gamma,m,r}<E^{\rm ssc}_\gamma \cr
}
\end{eqnarray}
}
and for slow cooling regime is
{\small
\begin{eqnarray}
\label{scsyn_rat}
F^{\rm ssc}_{\nu,r}\propto\cases{
 t^{-\frac{19}{30}}\,\left(E^{ssc}_{\rm \gamma,r}\right)^{\frac{1}{3}}  ,\hspace{1.3cm} E^{\rm ssc}_\gamma<E^{\rm ssc}_{\rm \gamma,m,r},\cr
t^{-\frac{5p+3/2}{5}}\,\left(E^{ssc}_{\rm \gamma,r}\right)^{-\frac{p-1}{2}},E^{\rm ssc}_{\rm \gamma,m,r}<E^{\rm ssc}_\gamma<E^{\rm ssc}_{\rm \gamma,c,r},\,\,\,\,\,\cr
t^\frac{-4p+1}{4}\,\left(E^{ssc}_{\rm \gamma,r}\right)^{-\frac{p}{2}},\,E^{\rm ssc}_{\rm \gamma,c,r}<E^{\rm ssc}_\gamma. \cr
}
\end{eqnarray}
}
Following  \cite{2003ApJ...597..455K},  the peak in the  SSC flux, for the fast-cooling regime,  can be written as 
{\small
\bary\label{ssc_peak}
F^{\rm ssc}_{\rm \gamma,peak}&\sim& F^{\rm ssc}_{\rm peak}    \epsilon_{B,r}^{-3/4}\,n^{-3/4}\,D^{-2}E^{3/2}\,\Gamma^{-5}\,T_{90}^{-5/2}\,{E^{\rm ssc}_{\rm \gamma,r}}^{-1/2},
\eary
}
with {\small $F^{\rm ssc}_{\rm peak}\sim  \frac{10^{-2}\, m_e^{5/2}\,(p-1)}{m_p^{11/4}\,q_e^{1/2}\,(p-2)}$}\,.
\section{Application: GRB 090510} 
GRB 090510 was detected on 2009 May 10 by both instruments aboard Fermi; Gamma-Ray Burst Monitor (GBM) and LAT \citep{2009GCN..9334....1O, 2009GCN..9336....1G} and  the three instruments aboard Swift; BAT, XRT and UVOT  \citep{2010ApJ...709L.146D,2009GCN..9337....1U, 2009GCNR..218....1H}.  The LAT instrument observed a bright peak at $0.3$ s followed by an extended  emission (above 100 MeV) during  200 s after the GBM trigger.  The extended component was described with a simple power law with a temporal decay index $\alpha_{LAT}=1.38\pm0.07$.\\
At 00:23:00 UT, 2009 May 10,  Swift BAT triggered on GRB 090510 \citep{2009GCNR..218....1H}. Refined analysis of BAT data found a duration of prompt emission T$_{90}=0.3\pm 0.1$ s \citep{2009GCN..9337....1U, 2010ApJ...709L.146D}.  Swift X-ray Telescope (XRT) started observing the X-ray afterglow of GRB 090510 at 98 s after the GBM trigger.  A broken power law was used to fit the LC. The best-fit parameters found were: an early decay slope $\alpha_{X,1}=0.74\pm0.03$, a break time t$_{br, X}=1.43^{+0.09}_{-0.15}\times 10^3$ s and a late decay slope $\alpha_{X,2}=2.18\pm0.10$.  The Swift  Ultra Violet and Optical Telescope (UVOT) began detecting this burst at 97 s after the initial trigger. This instrument measured the position of the optical afterglow counterpart to be  R.A. (J2000), decl. (J2000) = $22^h 14^m 12^s.5, -26^\circ34'59''.2$ \citep{2009GCN..9342....1K}. The optical LC was well fitted by a broken power law. In this optical band,  the best-fit parameters found were:  an early decay slope $\alpha_{opt,1}=-0.50^{+0.11}_{-0.13}$, a break time t$_{br, opt}=1.58^{+0.46}_{-0.37}\times 10^3$ s and a late decay slope $\alpha_{opt,2}=1.13^{+0.11}_{-0.13}$.   We summarize in Table 1 the relevant observation parameters for GRB 090510.\\ 
By considering the quantities inferred from observations given in Table 1,  the isotropic equivalent kinetic energy $E_{\rm k,iso}=E_{\rm \gamma,iso}/\eta= 10^{53} {\rm erg}$ with $\eta\approx 0.2$ the kinetic energy efficiency to convert bulk kinetic energy to $\gamma$-ray energy $E_{\rm \gamma,iso}$ and using the method of Chi-square minimization \citep{1997NIMPA.389...81B}, we obtain the values of density,  bulk Lorentz factors and the microphysical parameters that reproduce the multiwavelength (LAT, XRT and UVOT) afterglow observations  with the spectral index of electron distribution $p=2.2$ and $t_{dec}=$ 0.3 s.  The value of this spectral index was chosen linking  the observed slopes of temporal decays of LAT ($\alpha_{GeV}=1.125$; \cite{2011ApJ...733...22H}) and X-ray ($\alpha_{X,1}=0.74$ and  $\alpha_{X,2}=2.18$; \cite{2010ApJ...709L.146D}) fluxes with the synchrotron emission  LCs (eqs. \ref{fcsyn_t} and \ref{scsyn_t})   $\alpha_{LAT}=(3p-2)/4$, $\alpha_{X,1}=(3p-3)/4$ and  $\alpha_{X,2}=p$, respectively.   LAT detections are modeled by synchrotron emission from FS and SSC radiation from RS; the long-lasting flux at 100 s requiring synchrotron LC in the fast-cooling regime (eq. \ref{fcsyn_t})  and the bright peak flux at 0.3 s with SSC radiation,  for ultra-relativistic electrons radiating photons in the energy range of 0.1 - 4  GeV \citep{2010ApJ...709L.146D}.    The X-ray emission at $t < 1.43\, {\rm ks}$ is described using the synchrotron LC in the slow-cooling regime (eq. \ref{scsyn_t}) at  t= 100 s for electrons giving off at  $E^{syn}_{\gamma,f}=$ 1 keV, and  optical component at $t < 1.58\, {\rm ks}$ using the LC in slow-cooling regime (eq. \ref{scsyn_t})  at t=100 s for $E^{syn}_{\gamma,f} =$ 1 eV.   The values of microphysical  parameters, $\epsilon_{B,f}$, $\epsilon_{e}$ and the density in the range $10^{-6}\,\leq n \leq\,  10^{-1}$ ${\rm cm^{-3}}$, that describe the long-lasting components  of LAT, X-ray and optical data up to $\sim1.5 \times10^3$ s are plotted in Figure \ref{param_for} and the parameters $\epsilon_{B,r}$, $\epsilon_{e}$ and  $10^{-6}\, \leq n \leq\,  10^{-1}$ ${\rm cm^{-3}}$ that explain the LAT-peak data are plotted in Figure \ref{param_rev}.   After $\sim1.5 \times10^3$ s, we use the post jet-break equations for the LC, $F^{\rm jet}_{\nu,f}\propto t^{-p}$ for X-ray and  $\propto t^{-\frac{1}{3}}$ for optical flux \citep[e.g.][]{2010MNRAS.409..226K}. \\
\begin{center}\renewcommand{\arraystretch}{0.7}\addtolength{\tabcolsep}{-4pt}
\begin{center}
\scriptsize{\textbf{Table 1. Quantities inferred from the multiwavelength afterglow observation of GRB 090510.}} \\
\end{center}
\begin{tabular}{ l c c c c}
 \hline \hline
\scriptsize{\bf{GeV flux}} & \\ 
\hline \hline
\scriptsize{Power index}	&  \scriptsize{$\alpha_{GeV}$}	 	&		\scriptsize{$1.38\pm0.07 $}	\\
\scriptsize{Initial time after the GBM trigger (s)}&\scriptsize{$t_{0,GeV}$ }	 	&		\scriptsize{$ 0.013$}		\\
\scriptsize{Duration of extended emission (s)} &\scriptsize{$t_{ee, GeV}$ }	 	&		\scriptsize{$ \sim 200$}	\\
\scriptsize{Duration of bright peak (s)}&\scriptsize{$t_{bp,GeV}$ }	 	&		\scriptsize{$< 0.3 $}		\\
\hline \hline
\scriptsize{\bf{X-ray flux}}		         & 		                                                                                  \\
\hline \hline
\scriptsize{Early decay slope}&\scriptsize{$\alpha_{X,1}$}			 &			\scriptsize{$0.74\pm0.03$}	 \\
\scriptsize{Late decay slope}&\scriptsize{$\alpha_{X,2}$}			 &			\scriptsize{$2.18\pm0.10$}	\\
\scriptsize{Break time (s)}&\scriptsize{$t_{br,X}$ }			 &		\scriptsize{$1.43^{+0.09}_{-0.15}\times 10^3$} 	\\
\scriptsize{Initial time after the GBM trigger (s)}&\scriptsize{$t_{0,X}$ }	 	&		\scriptsize{$98$}		\\
\hline\hline
\scriptsize{\bf{Optical flux}}  		         &                                                                                  \\
\hline\hline
\scriptsize{Early decay slope} &\scriptsize{$\alpha_{opt,1}$}		&		\scriptsize{$0.50^{+0.11}_{-0.13} $}		\\
\scriptsize{Late decay slope}&\scriptsize{$\alpha_{opt,2}$}		&		\scriptsize{$1.13^{+0.11}_{-0.13}$}			\\
\scriptsize{Break time (s)}&\scriptsize{$t_{br,opt}$ }			&			\scriptsize{$1.58^{+0.46}_{-0.37}\times 10^3$}		\\
\scriptsize{Initial time after the GBM trigger (s)} &\scriptsize{$t_{0,opt}$ }	 	&		\scriptsize{$97$}					 \\
\hline
\scriptsize{Isotropic energy (erg)}&\scriptsize{$E_{\rm iso}$ }					& 		\scriptsize{$(1.08\pm 0.06)\times 10^{53}$}	\\
\scriptsize{Duration of prompt emission (s)}&\scriptsize{$T_{90}$ }				& 		\scriptsize{$0.3 \pm 0.1$}		\\
\scriptsize{Redshift}&\scriptsize{$z$}					& 		\scriptsize{$0.903\pm0.001$}	\\
\hline
\end{tabular}
\end{center}
\begin{flushleft}
\scriptsize{
\textbf{References}. (1) \cite{2010ApJ...716.1178A}; (2) \cite{2009GCN..9353....1R}; (3) \cite{2010A&A...516A..71M}; (4) \cite{2009GCN..9337....1U}; \cite{2010ApJ...709L.146D}.} 
\end{flushleft}
Fig. \ref{param_for} displays regions in orange, purple and green colors.  The zones in orange colors  show  the set of parameters that describes the long-lasting LAT component (in the energy range of  0.1 - 4  GeV at 100 s), in purple colors present those parameters that describe the X-ray emission for $t< 1.43\,{\rm ks}$ (at 1 keV and 100 s) and in green ones those parameters that explain the optical component for $t< 1.58\,{\rm ks}$ (at 1 eV and 100 s).  The black lines over the area in orange color represent the photon fluxes emitted at 1 GeV, 500 MeV, 200 MeV and 100 MeV.  The zones intersected (orange, purple and green colors) correspond to the set of parameters that describe more than one component. Then,  it is possible to find a set of parameters (intersected zones) that  explain the LAT, X-ray and optical components for the range of density values considered.  For instance, in the panel with $n=10^{-1}\,{\rm cm^{-3}}$ label, the long-lasting LAT, X-ray and optical components are described by the microphysical parameters around $\epsilon_{e}\sim$ 0.44 and  $\epsilon_{B,f}\,\sim10^{-5.3}$ for ultra-relativistic electrons giving off at $\sim$ 1 GeV. As shown in fig.  \ref{param_for},  these surfaces are shifted to smaller values of $\epsilon_{e}$ and higher values of $\epsilon_{B,f}$ as  density decreases.  As can be seen, the LAT component  does not give a constrain on $\epsilon_B$, since the orange zone is very ``flat".  This is due to  the LAT energy band is above the cooling energy, and thus, the flux is almost independent on $\epsilon_B$ \citep{2000ApJ...538L.125K}. Similarly, due to the synchrotron emission with $E^{syn}_{\rm \gamma, f}>E^{syn}_{\rm \gamma, c, f}$ does not depend on the density,  the orange region does not change when the density varies \citep[e.g.][]{2000ApJ...538L.125K}.    \\
Figure \ref{param_rev} exhibits regions in orange color. Each of these regions show the set of microphysical parameters ($\epsilon_{B,r}$ and $\epsilon_{e}$)  for the density in the range of $10^{-6}\,\leq n \leq\,  10^{-1}\, {\rm cm^{-3}}$ that explain the bright LAT-peak component. The black lines over the region in orange color represent the photon emissions radiated at 1 GeV, 500 MeV, 200 MeV and 100 MeV. The value of $\epsilon_e$ for each density obtained as a good fit to the long-lasting LAT, X-ray and optical data (see fig. \ref{param_for}) is highlighted in a dashed line.   To constrain the pair of values of $\epsilon_e$ and $n$ that describe the bright LAT-peak data,  we plot the break photon energy of SSC emission from RS as a function of magnetic field equipartition, as shown Figure \ref{Ebreak}. This figure displays that the break SSC energy lies in the energy range of LAT instrument only for n=$10^{-1}\,{\rm cm^{-3}}$ and $\epsilon_{B,r}\gtrsim 0.01$.  Comparing figs. \ref{param_rev} and \ref{Ebreak} can be seen that for  $n=10^{-1}\,{\rm cm^{-3}}$, $\epsilon_e=0.43$ and $0.05\lesssim\epsilon_{B,r}\lesssim0.3$,  ultra-relativistic electrons radiating photons at $E^{ssc}_{\gamma,m,r}\sim 100$ MeV would describes the bright peak in the LAT data.  For this case,  the SSC radiation from RS around the peak ($E^{ssc}_{\gamma,r}=E^{ssc}_{\gamma,m,r}\sim 100$ MeV),  would have a rise flux  $F^{ssc}_\nu \propto t^\frac{p-2}{4}$ and temporal decay  $F^{ssc}_\nu \propto  t^{-\frac{3}{4}}$ (eqs. \ref{fcsyn_rbt} and \ref{fcsyn_rat}).\\
With the value of density $n=10^{-1}\,{\rm cm^{-3}}$, we found that  the bulk Lorentz factor is $\Gamma_r\simeq 3000 > \Gamma_c$ and the RS evolves in the thick-shell case with  a critical Lorentz factor $\Gamma_c=1300$.  When the bulk Lorentz factor drops below the inverse of the jet opening angle $\theta_j\simeq 0.72^\circ$ \citep{2010MNRAS.409..226K},  we see a break in the LC \citep{1999ApJ...519L..17S}, which corresponds to a jet-break time of $t_{\rm jet}\simeq 1.5 \times 10^3 {\rm s}$.\\
In Table 2, we sum up the microphysical parameters, densities and bulk Lorentz factors found after modeling the multiwavelength afterglow observed in GRB 090510, as shown in Figure \ref{fit_afterglow}. This figure displays  all contributions to the multiwavelength afterglow observations of GRB 090510.  From the value of  magnetic microphysical parameter of RS, the magnetization parameter is $\sigma\simeq$ 0.3.  This result agrees with the observed bright peak in the LAT flux.  If the GRB outflow would have been highly magnetized ($\sigma\gg$1) when it crossed the RS,  particle acceleration may be very inefficient, as pointed out by \cite{2011ApJ...726...75S} and the RS would have been suppressed \citep{2005ApJ...628..315Z}.  Therefore a moderate  magnetization ($\sigma\leq1$) is needed (as found in this work) to obtain a bright peak from RS \citep{2003ApJ...595..950Z,2003MNRAS.346..905K,2004A&A...424..477F}.  Similarly,  comparing the microphysical parameter for the magnetic field in both shocks, it is possible observe that magnetic field in both shocks are different $B'_r=\mathcal{R}^{-1/2}_B \,B'_f\simeq200\,B'_f$, thus indicating that the ejecta carries a significant magnetic field.\\ 
%
%
\begin{center}
\begin{center}
\scriptsize{\textbf{Table 2. Values of microphysical parameters, densities and bulk Lorentz factors found for GRB 090510. }}
\end{center}
\begin{tabular}{ l c c c c c}
 \hline
 \hline
 FS& & & &RS \\
\hline
\scriptsize{$\epsilon_{B,f}$}    & \scriptsize{ $3\times 10^{-6}$ } & &  &\scriptsize{$\epsilon_{B,r}$}    & \scriptsize{ 0.28} \\
\scriptsize{$\epsilon_{e,f}$}    & \scriptsize{0.4}                        & & &\scriptsize{$\epsilon_{e,r}$}    & \scriptsize{0.4}  \\
\scriptsize{$n\,\, (\rm cm^{-3})$}    & \scriptsize{$10^{-1}$}  & & &\scriptsize{$n\,\, (\rm cm^{-3})$}   & \scriptsize{$10^{-1}$}\\
\scriptsize{$\Gamma_f$}    & \scriptsize{ 520} &  & &\scriptsize{$\Gamma_r$}    & \scriptsize{$3\times 10^3$}   \\
 \hline
\end{tabular}
\end{center}
\begin{center}
\end{center}
Based on the shock jump conditions \citep{2005ApJ...628..315Z} with the values found of the magnetization parameter $\sigma\simeq0.3$ and the FS bulk Lorentz factor $\Gamma_f=\gamma_2\simeq520$, then the relative Lorentz factor of the reverse shocks upstream and downstream and  the initial Lorentz factor around the discontinuity  are $\gamma_{34}\simeq 500$ and  $\gamma_4\simeq5\times10^5$, respectively.   The high value obtained of $\gamma_4$  is due to the fact that similar densities $n_1=n_4$ were considered \citep{2005ApJ...628..315Z}.  If we would  have allowed for a different ratio of densities, then this value would have been smaller.  The value found for similar densities could be explained in the context of the dynamical evolution of a hybrid relativistic outflow with arbitrary magnetization \citep{2015ApJ...801..103G}. In addition, the value of a magnetization parameter $\sigma \sim 0.3$ at the deceleration radius indicates that before deceleration the jet must also have dissipated a significant amount of Poynting flux during the prompt emission phase. A candidate process to make this happen would be the so-called internal collision-induced magnetic reconnection and turbulence model proposed by \cite{2011ApJ...726...90Z}.\\
%
%
Using the values of parameters reported in Table 2,  we have derived the observable quantities, as shown in Table 3. The values of the critical Lorentz factor computed in our model is self-consistent with the fact that the bright LAT peak  occurs at the end of the prompt phase and the RS evolves in the thick-shell case ($\Gamma_r> \Gamma_c$).\\
 The synchrotron self-absorption energies  from FS and RS are in the weak self-absorption regime, then, as observed in LC of GRB 090510, there is no thermal peak in the synchrotron spectrum \citep{2004ApJ...601L..13K,2013MNRAS.435.2520G}. \\
Due to the early LAT component is present between 0.1 - 4 {\rm GeV} and the break energy at the KN regime is 12.11\, {\rm GeV},  the LC of RS SSC derived here (eq. \ref{fcsyn_rat}) and used to describe the bright LAT peak is not altered.    Although we describe the long-lasting LAT component in the energy range of 0.1 - 4 GeV with synchrotron radiation from FS, the 12 events with energies $>$ 1 GeV and the most energetic event with $30.5^{+5.8}_{-2.6}$ GeV  at 0.829 s, we can not discard that these photons can have their origin in SSC emission from FS \citep[e.g.][]{2013ApJ...771L..33W,  2015MNRAS.454.1073B}.   It is worth noting that bursts with sub-TeV photons at dozens of seconds from the trigger might be interpreted as SSC radiation from FS and be detected by TeV $\gamma$-ray observatories as the High Altitude Water Cherenkov observatory (HAWC) \citep{2012APh....35..641A,2014arXiv1410.1536A}.
%
%
\begin{center}
\begin{center}
\scriptsize{\textbf{Table 3. Observables derived for  GRB 090510}}\\
\end{center}
\begin{tabular}{ l c c c c c}
   \hline
 \hline
 FS & & & & RS \\
\hline\hline
\scriptsize{$t_{dec}$ (s)} & \scriptsize{$0.3$} & & & \scriptsize{$\Gamma_c$} & \scriptsize{$1.3\times 10^3$}  \\
\scriptsize{$t_{jet}$ (s)} & \scriptsize{$1.5\times10^3$} & & &\scriptsize{$B_r$ (G)} &  \scriptsize{$610.1$} \\
\scriptsize{$B_f$ (G)} &\scriptsize{$2.9$} & & &\\
\hline
\small{Synchrotron}\\
\hline
\scriptsize{$E^{syn}_{\rm \gamma,a,f}$ (eV)}    & \scriptsize{$1.2\times 10^{-6}$} & & &\scriptsize{$E^{syn}_{\rm \gamma,a,r}$ (eV)}    & \scriptsize{$4.3\times 10^{-11}$} \\
\scriptsize{$E^{syn}_{\rm \gamma,m,f}$ (keV)}    & \scriptsize{$6.1$}  & & &\scriptsize{$E^{syn}_{\rm \gamma,m,r}$ (keV)}    & \scriptsize{$0.2$} \\
\scriptsize{$E^{syn}_{\rm \gamma,c,f} $} (GeV)   & \scriptsize{0.9}  & & &\scriptsize{$E^{syn}_{\rm \gamma,c,r}$ (eV)}    & \scriptsize{$10.4$} \\
\scriptsize{$E^{syn}_{\rm \gamma,max,f}$ (GeV)}    & \scriptsize{$86.4$}& &  &\\
\hline
\small{SSC}\\
\hline
\scriptsize{$E^{ssc}_{\rm \gamma,m,f}$ (TeV)}    & \scriptsize{$96.3$}  & & &\scriptsize{$E^{ssc}_{\rm \gamma,m,r}$ (MeV)}    & \scriptsize{$101.5$} \\
\scriptsize{$E^{ssc}_{\rm \gamma,c,f}$ (TeV)}    & \scriptsize{$18.9$}  & & &\scriptsize{$E^{ssc}_{\rm \gamma,c,r}$ (eV)}    & \scriptsize{$5.6$} \\
\scriptsize{$E^{KN}_{\rm \gamma,f}$ (TeV)}    & \scriptsize{$27.1$}  & & &\scriptsize{$E^{KN}_{\rm \gamma,r}$ (GeV)}    & \scriptsize{$12.1$}  \\
\hline
\end{tabular}
\end{center}
%
%
%
\section{Conclusions}
\begin{figure}
\epsscale{1.2}
\plotone{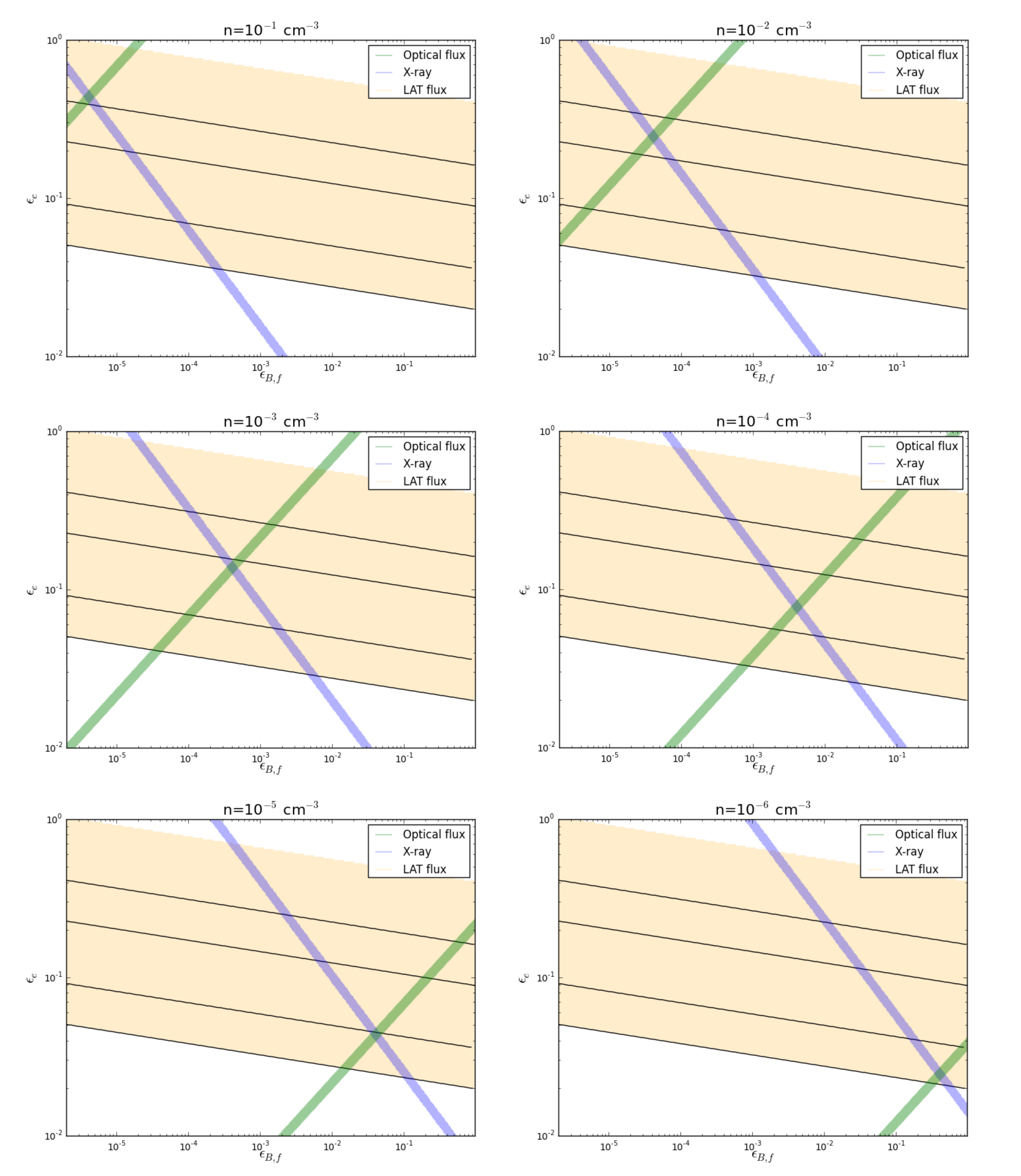}
\caption{Values of microphysical parameters ($\epsilon_{B,f}$ and $\epsilon_{e}$) obtained through synchrotron emission from FS that describe the long-lasting LAT, X-ray and optical emission before the break time at $\sim\,1.5\times 10^3\, {\rm s}$.}
\label{param_for}
\end{figure}

\begin{figure}
\epsscale{1.2}
\plotone{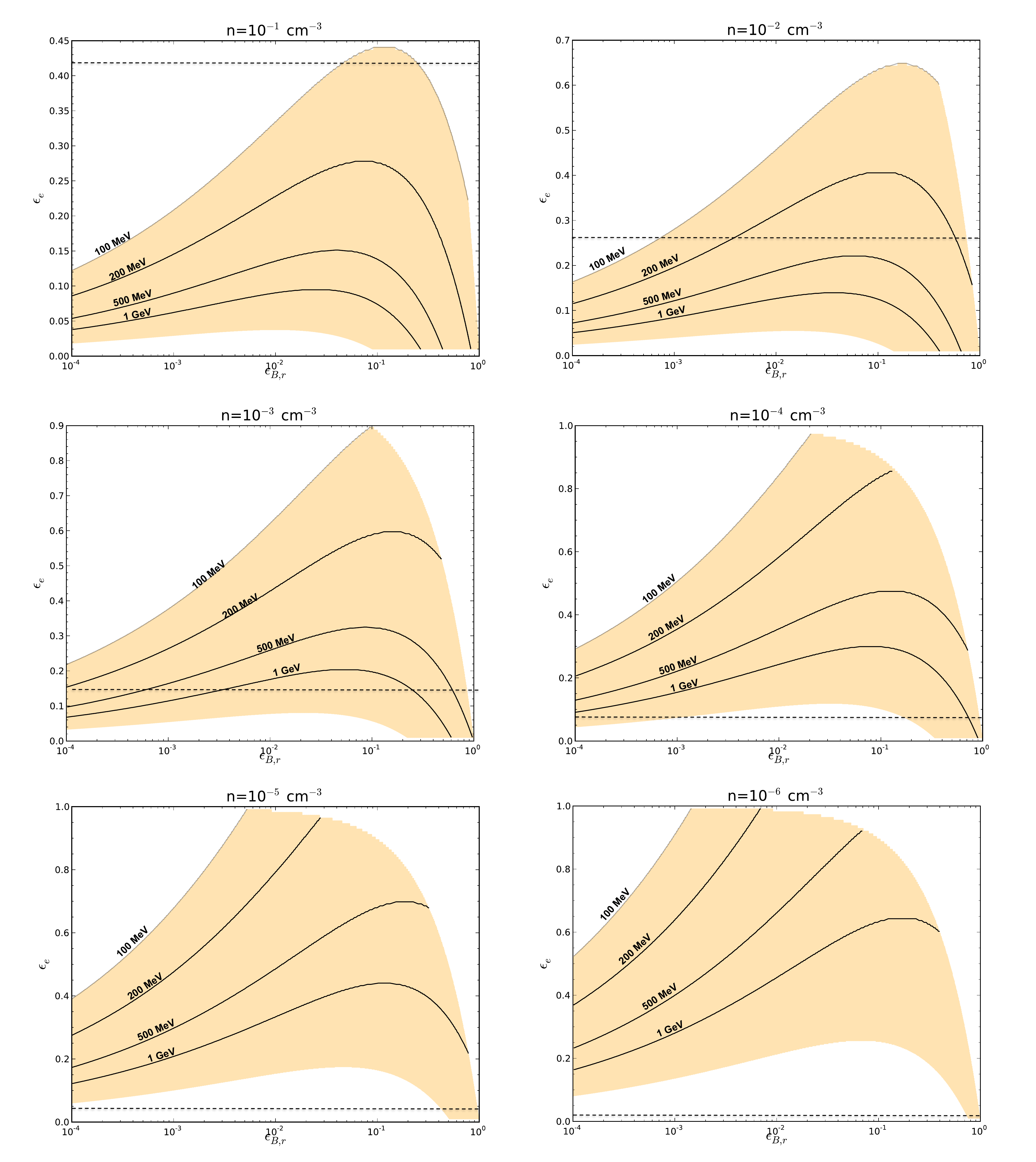}
\caption{Values of microphysical parameters ($\epsilon_{B,r}$ and $\epsilon_{e}$) obtained through SSC emission from RS that describe the bright LAT-peak flux.   The value of $\epsilon_e$ for each density obtained as a good fit to the extended LAT, X-ray and optical fluxes of FS (see fig. \ref{param_for}) is highlighted in a dashed line.}
\label{param_rev}
\end{figure}

\begin{figure}
\epsscale{.65}
\plotone{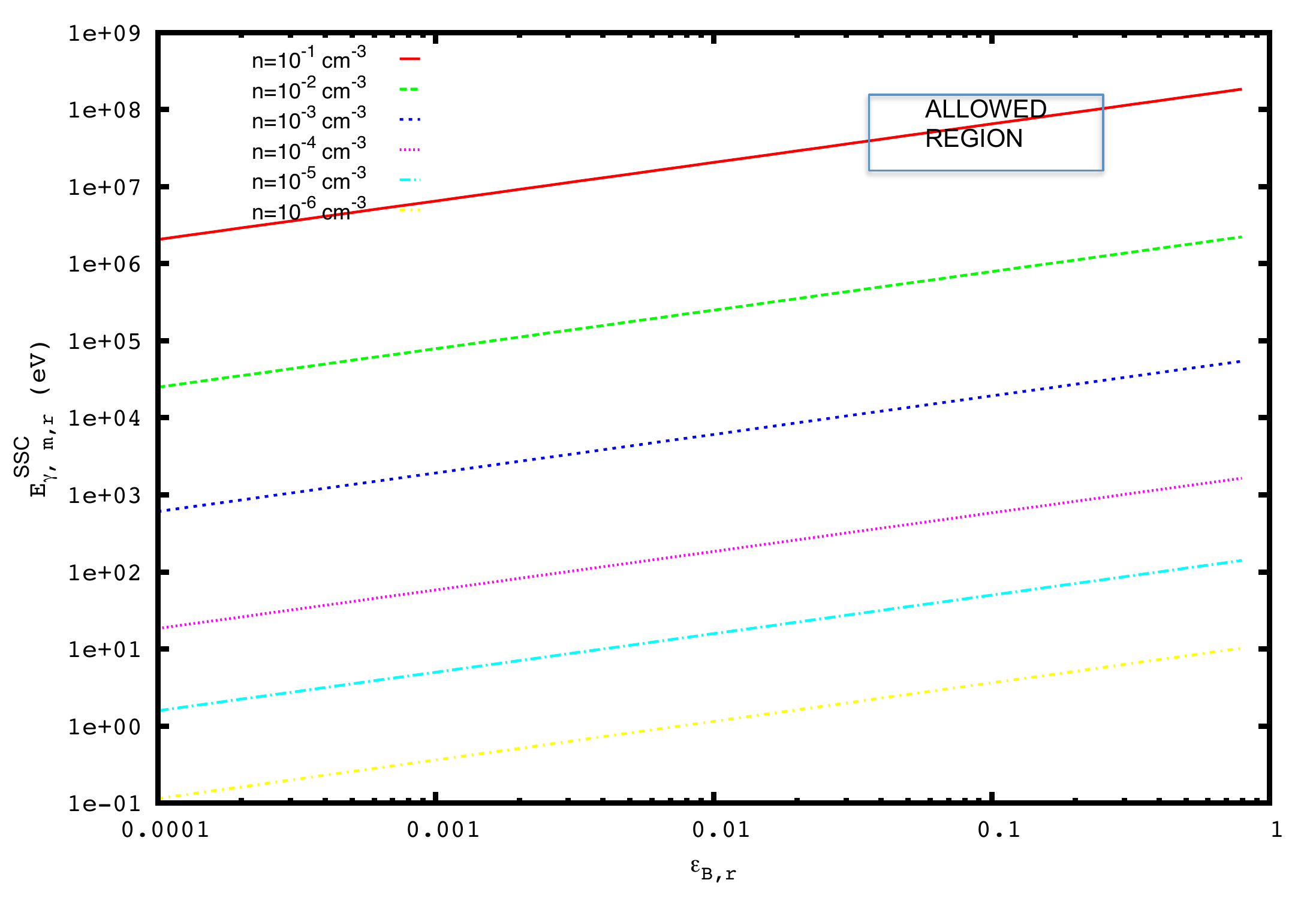}
\caption{Break photon energy of SSC emission from reverse shock as a function of magnetic equipartition parameter.  We consider the values of density and electron equipartition parameter obtained in Figures  \ref{param_for} and \ref{param_rev}.}
\label{Ebreak}
\end{figure}

\begin{figure}
\epsscale{.65}
\plotone{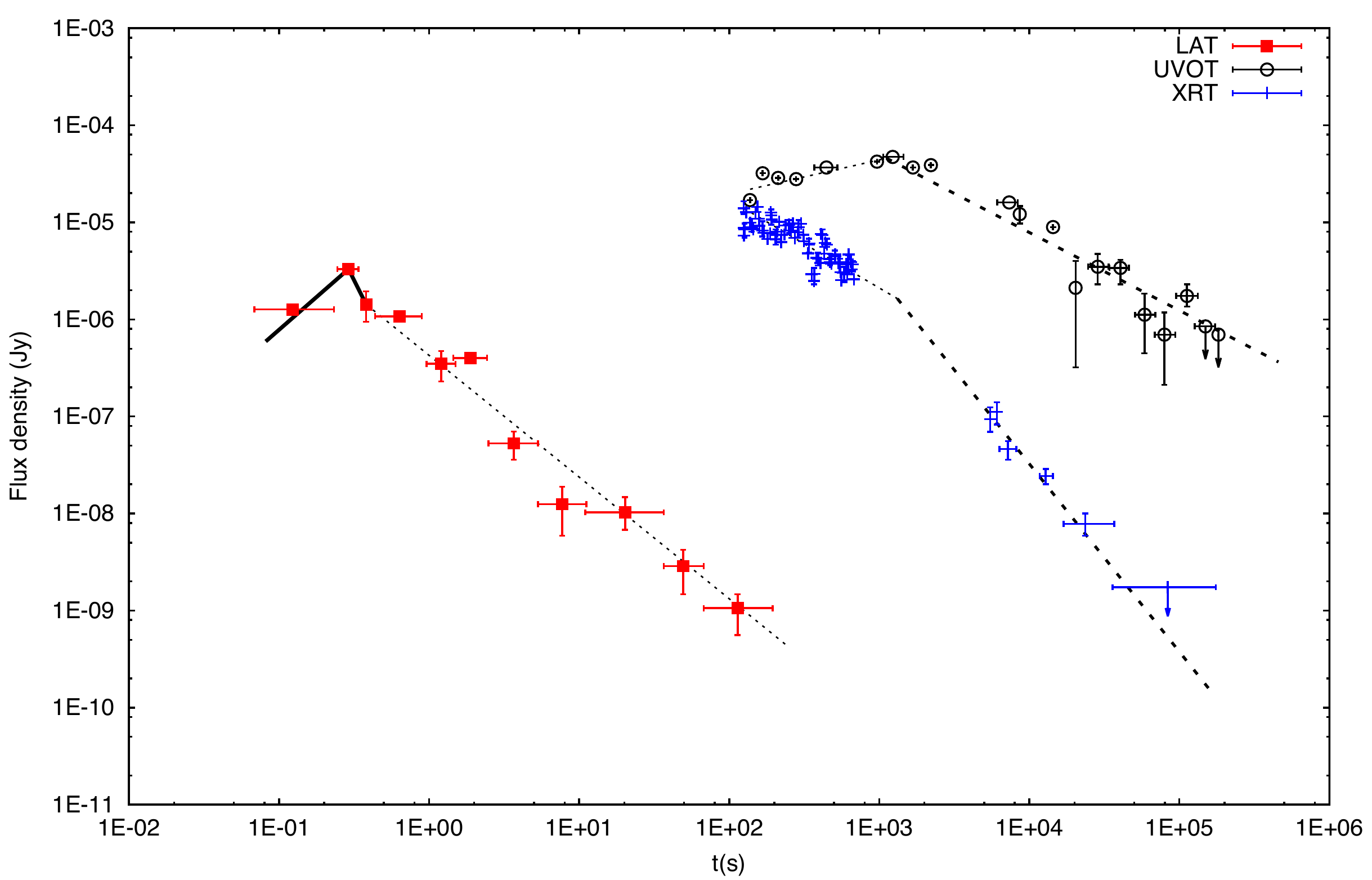}
\caption{Fits of the multiwavelength (LAT, XRT and UVOT) observation of GRB 090510.  We use the LC of  RS  in the thick-shell regime to describe the bright LAT-peak flux (continuous line), the LC of FS to explain  the temporally extended LAT, X-ray and optical emissions  before the break time at $t_{br}\sim\,1.5\times 10^3\, {\rm s}$  (dotted lines) and the LC  after the jet break time (dashed lines).}
\label{fit_afterglow}
\end{figure}

We have introduced an external shock model to explain the multiwavelength afterglow observations present in GRB 090510. Taking into account that the LAT-peak flux occurs at the end of the prompt emission, we have considered that the ejecta propagating in the homogeneous medium is decelerated early at $\sim 0.3$ s and the RS evolves in the thick shell regime.\\
Under the standard assumptions that the magnetic field and electron microphysical parameters are constant,  we have modeled the long-lasting  LAT, X-ray and optical components up to the break time ($t_{br}\sim 1.5\times 10^3$ s)  with the synchrotron radiation LC from FS and after this break time, the LC of spreading jet has been used to fit the X-ray and optical fluxes.  The bright LAT-peak component has been described through LCs of SSC emission from RS (eqs. \ref{fcsyn_rbt} and \ref{fcsyn_rat}).  We  have plotted the set of values of density and microphysical parameters that describes these observations, as shown in figs.  \ref{param_for},  \ref{param_rev} and \ref{Ebreak}.    Considering the number density in the range   $10^{-6}\,\leq\, n\, \leq\,10^{-1}\,{\rm cm^{-3}}$ the microphysical parameters lie in the range $10^{-5.5}\leq\epsilon_{B,f}\leq 0.7$ and $10^{-1.5}\leq\epsilon_{e}\leq 0.48$ for FS and in the RS, the microphysical parameters lie in the range $10^{-4.0}\leq\epsilon_{B,r}\leq 0.9$ and $10^{-1.6}\leq\epsilon_{e}\leq 0.45$. From RS, one can see that  density values lower than $n\leq 10^{-4}\,{\rm cm^{-3}}$ require SSC energies higher than 1 GeV ($\epsilon^{ssc}_{\gamma,r}\geq$ 1 GeV).  In this case,  it is not possible with plausible parameter values to reach these SSC energies.  Hence,  the standard RS cannot contribute to the LAT emission, thus reproducing the results obtained by \cite{2011ApJ...733...22H}.  Otherwise, requiring the temporal and spectral description of the bright LAT-peak flux, we get the number density $n =  10^{-1}\,{\rm cm^{-3}}$ and the values of parameters $\epsilon_{e}= 0.44$, $\epsilon_{B,f}= 10^{-5.5}$ and  $\epsilon_{B,r}= 0.24$. The observed X-ray and UVOT flux decay indices after the $t_{bk}$ are $\alpha_{X,2}=2.18\pm0.10$ and $\alpha_{opt,2}=1.13^{+0.11}_{-0.13}$, respectively, which are softer than the electron synchrotron radiation. These temporal decays are more consistent with the evolution of the jet after it slows down and spreads for X-ray  rather than optical data.  The bulk Lorentz factor obtained at the $t_{br}$ is $\sim$ 120 that corresponds to a jet opening angle of $0.72^{\circ}$. \\
 It is worth noting that in the current model the differences between the values of SSC and synchrotron spectral breaks at FS and RS comes from the magnetic energy fractions given at both shocks.   Comparing the strength of magnetic fields at FS ($\sim 3\, {\rm G}$) and RS ($\sim 600\, {\rm G}$), we can see that the magnetic field in the RS region is stronger ($\sim$ 200 times) than in the FS region which indicates that the ejecta is magnetized.  The magnetization of the ejecta  modifies the temporal characteristics of the bright LAT-peak component; it becomes much shorter than $\leq 0.3$ s \citep{2004A&A...424..477F}.  Therefore, the short and bright LAT-peak component at the end of the prompt emission forecasts the mechanism of jet production to be due to magnetic processes instead of neutrino annihilations, thus giving evidence of the magnetic field amplification in neutron star mergers  in GRB 090510.   In principle, these two mechanisms (neutrino annihilation and magnetic processes) could supply the outflow with different energies.  The outflows driven by neutrino annihilation can provide $10^{48}\, {\rm erg}$ as shown in GRB 050509B \citep{2005ApJ...630L.165L} whereas the magnetic mechanism could supply $10^{51}\, {\rm erg}$ or more \citep{2003MNRAS.345.1077R, 1992ApJ...395L..83N} as in the case for this burst.  Additionally,  based on compact object binary population synthesis models, \cite{2002ApJ...570..252P} argued that such merger tends to happen in lower density environments $\sim 0.1\, {\rm cm^{-3}}$  which is in accordance with our results.\\
Some authors have suggested that the fireball wind which is connected to the GRB central engine, a black hole (BH) - torus system or a rapidly rotating magnetar may be endowed with ``primordial" magnetic fields\citep{1992Natur.357..472U, 2000ApJ...537..810W, 1997ApJ...482L..29M}.  The energy requirement (isotropic-equivalent luminosities $L_{\gamma, iso}\geq 10^{52} {\rm erg/s}$)  demands magnetic fields at the base in excess of B$\sim 10^{15}\,{\rm G}$. In the LAT era,  GRB110731A \citep{2013ApJ...763...71A} and GRB130427A \citep{2014Sci...343...42A}  were detected from optical to GeV energy range and showed to have in the LAT LC temporally extended fluxes lasting hundreds of seconds in coincidence with  short-lasting bright peaks.  \cite{2015ApJ...804..105F} and \cite{2016ApJ...818..190F} showed that both components could be interpreted as synchrotron and SSC emissions from the forward and reverse shocks, respectively, provided that the central engines were entrained with a significant magnetic field. Authors found that the strength of magnetic fields in the reverse-shock region were stronger ($\sim$ 50 and 66 times, respectively) than in the forward-shock region. Comparing the isotropic total energies and the values found of magnetic field in the RS for GRB110731A, GRB130427A and GRB090510, we can see that GRB090510 demands more magnetic fields at the base of the jet. For instance, considering a typical size of fireball for sGRBs ($r_i\sim 10^{6.5}{\rm cm}$;  \cite{2004ApJ...608L...5L,2005ApJ...632..421L, 2007PhR...442..166N}) and total isotropic energy $10^{53} {\rm erg}$, the magnetic field at the initial time is rough calculated as $B\approx\sqrt{8\epsilon_{B,i}E_{\gamma,iso}/r^3_i}\approx \sqrt{8 \epsilon_{B,i} 10^{32}}\, {\rm G}$ which is much stronger than the typical magnetic field of the neutron star $\sim 10^{12}\, {\rm G}$. Here, $\epsilon_{B,i}$ is the initial fraction of total energy given to magnetic field.  We have shown that the presence not only of GRB 090510 but also in future detections of sGRBs  with the same features could offer evidence of the magnetic field amplification during the merger of NS- NS.\\
The main difference between the current model and the model proposed by \cite{2011ApJ...733...22H} is that the RS evolves in the thick-shell case and the ejecta must be magnetized in order to obtain a good description of the bright LAT-peak flux.  We emphasize that in the external shock model of  early afterglow with the suitable values of parameters obtained in this work, the magnetized RS SSC emission can explain successfully the early LAT data.  Recently,  \cite{2015ApJ...810..160G} performed a morphological analysis of the early optical light curves in some GRBs including GRB 090510 \citep{2010arXiv1002.2863P}. As successful descriptions of optical light curves, they found values of  microphysical parameters similar to those obtained in this work, indicating that the jet was magnetized.\\
Using the MeV prompt emission (GBM) data and assuming a constant radiative efficiency, \cite{2011MNRAS.415...77M} were able to track the energy accumulation in the external shock with an internal/external shell model code.  Authors analyzed some LAT bursts (including GRB 090510) and suggested  that the high-energy emission present in most of LAT burst during the prompt phase is most likely a superposition of a gradually enhancing external shock component and a dominant emission component that is of an internal origin.  Similarly,  \citet{2011ApJ...730..141Z} performed a comprehensive analysis of Fermi GRB data to study the possible origins of LAT/GBM GRBs. For this analysis, they derived LAT/GBM LC for GRB090510 (see fig. 8 in \citealp{2011ApJ...730..141Z}). The  lowest panel shows that GeV emission peaks at an epoch when the MeV emission has already decayed. Therefore, at the GeV emission peak the external shock is likely not undergoing energy injection and has entered the deceleration phase. Additionally, LAT/GBM LCs of some bursts derived in \citet{2011ApJ...730..141Z} exhibit coincidences between peaks. For instance, the GeV peak present in GRB 080916C coincides with the second bright peak in the GBM LC, suggesting that GeV emission is the spectral extension of MeV emission to higher energies \citep{2011ApJ...730..141Z}.  However,  other bursts such as GRB 090510 do not exhibit these remarkable coincidences between peaks  (see fig. 8,  \citealp{2011ApJ...730..141Z}).  Finally,  the author concluded that for the LAT/GBM bursts analyzed (all except GRBs 090902B and 090510), the LAT and GBM photons consistently belong the same spectral component, suggesting a possible common origin. \\
\acknowledgments
We thank the anonymous referee  for a critical reading of the paper and valuable suggestions that helped improve the quality and clarity of this work.  We thank Peter Meszaros, Anatoly Spitkovsky, Dimitrios Giannios and Fabio de Colle  for useful discussions. This work was supported by the project PAPIIT  IG100414. PV thanks Fermi grant
NNM11AA01A and partial support from OTKA NN 111016 grant. 
%
%
%

%
%
\clearpage
\clearpage
\clearpage
\clearpage

\appendix
The quantities are given by 
{\small
\bary\nonumber
A^{syn}_{\rm fl}&\simeq& \frac{m_e^{2/3}\,\sigma_T^{5/3}}{3^{3/2}\,2^{1/2}\,\pi^{5/6} \,q_e^{4/3}\,m_p^{1/6}}\, (1+Y_f)^{\frac23}\,(1+z)^\frac{7}{6}\, \epsilon_{B,f}\,n^\frac56\,D^{-2}\,E^\frac{7}{6}\,,\cr
A^{syn}_{\rm fm}&\simeq& \frac{m_e^{3/2}}{3^{1/4}\,8 \,q_e^{1/2}\,m_p} \,(1+Y_f)^{-1}  \,(1+z)^\frac{3}{4}\, \epsilon_{B,f}^{-\frac{1}{4}}\,D^{-2}\,E^\frac{3}{4}\,,\cr
A^{syn}_{\rm fh} &\simeq& \frac{m^3_e\,(p-1)}{3^{1/2}\,8\,q_e\,m_p^2\,(p-2)}  \left(\frac{3^{1/2}\,q_e\,m_p^2\,(p-2)^2}{m_e^3\,(p-1)^2} \right)^\frac{p}{2}\, (1+Y_f)^{-1} (1+z)^\frac{2+p}{4}\,\epsilon_{B,f}^\frac{-2+p}{4}\,\epsilon_{e}^{-1+p}\,D^{-2}\,E^\frac{2+p}{4}\,, \cr
A^{\rm syn}_{\rm sl}&\simeq&\frac{m_e^2\,\sigma_T (p-1)^{2/3}}{3^{7/6}\,2^{3/2}\,\pi^{1/2} \,q_e^{4/3}\,m_p^{7/6}\,(p-2)^{2/3}} \,(1+z)^\frac{5}{6}\,\epsilon_{B,f}^\frac13\,\epsilon_{e}^{-\frac{2}{3}}\,n^\frac12\, D^{-2}\,E^\frac{5}{6}\,,\cr
A^{\rm syn}_{\rm sm}&\simeq&\frac{m_e\,\sigma_T}{6 (2\pi)^{1/2}\,q_e\,m_p^{1/2}}  \left(\frac{3^{1/2}\,q_e\,m_p^2\,(p-2)^2}{m_e^3\,(p-1)^2} \right)^\frac{p-1}{2} (1+z)^\frac{p+3}{4}\,\epsilon_{B,f}^\frac{p+1}{4}\,\epsilon_{e}^{p-1}\,n^\frac12\,D^{-2}\,E^\frac{p+3}{4}\,,\cr
\eary
}
where $A^{syn}_{\rm sh}=A^{syn}_{\rm fh}$.
\end{document}